\documentclass[preprint,showpacs,preprintnumbers,amsmath,amssymb]{revtex4}

\usepackage{graphicx}
\usepackage{dcolumn}
\usepackage{bm}

\newcounter{sub}
\newcounter{subeqn}[sub]
\def\be{\begin{equation}}
\def\ee{\end{equation}}

\def\st{\stepcounter{sub}}

\def\bea{\begin{eqnarray}}
\def\eea{\end{eqnarray}}

\newcommand\eeta{{\mbox{\boldmath $\eta$}}}
\newcommand\etas{{\mbox{\scriptsize{$\eeta$}}}}
\newcommand\nab{\mbox{\boldmath $\nabla$}}

\def\k{{\bf k}}
\def\l{{\bf l}}
\def\m{{\bf m}}

\def\r{{\bf r}}
\def\v{{\bf v}}

\newcommand\no{\nonumber}

\newcommand\GTP{{\rm GTP}}
\newcommand\GDP{{\rm GDP}}
\newcommand\Tu{{\rm T}}
\newcommand\MT{{\rm MT}}

\begin{document}


\title{A first principle (3+1) dimensional model for microtubule polymerization 
}

\author{Vahid Rezania$^{1,2}$ and Jack A. Tuszynski$^1$}
\address{1-Department of Physics, University of Alberta, Edmonton, AB T6G 2J1, Canada
\\
2 - Department of Science, Grant MacEwan College, Edmonton, AB T5J 2P2, Canada}


\begin{abstract}
In this paper we propose a microscopic model to study the polymerization of microtubules (MTs).
Starting from fundamental reactions during MT's assembly and disassembly processes,
we systematically derive a nonlinear system of equations that determines the dynamics of microtubules in 3D.
We found that the dynamics of a MT is mathematically
expressed via a cubic-quintic nonlinear Schr\"{o}dinger (NLS) equation.
Interestingly, the generic 3D solution of the NLS equation
exhibits linear growing and shortening in time as well as temporal fluctuations about
a mean value which are qualitatively similar to the dynamic instability of MTs observed experimentally.
By solving equations numerically, we have found
spatio-temporal patterns consistent with experimental observations.
\end{abstract}


\maketitle

\section{Introduction}\label{int}

More than 25 years ago, Del Giudice et al. \cite{Del85} argued that
a quantum field theory approach to the collective behaviors of biological
systems is not only applicable but the most adequate as it leads naturally to
nonlinear, emergent behavior which is characteristic of biological organization.
They followed a line of reasoning championed by Davydov \cite{Dav82,Dav79} and Fr\"ohlich \cite{Fro77}
who emphasized integration of both conservative and dissipative  mechanisms in biological
matter leading to the emergence of spatio-temporal coherence with various specific
manifestations such as almost lossless energy transport and long-range coordination.

Microtubules (MTs) are long protein polymers present in almost all living cells.
They participate in a host of cellular functions with great specificity and
spatio-temporal organization.
Microtubules are assembled by tubulin polymerization into a helical
lattice forming a cylinder which is rigid and straight by biological standards.
These protein polymers, typically several microns long, participate in
fundamental cellular processes such as locomotion, morphogenesis, and reproduction \cite{ALRW94}.
It is also suggested that MTs are responsible for transferring mechanical energy across
the cell, with little or no dissipation.

Both \emph{in vivo} and \emph{in vitro} observations confirmed that
an individual microtubule switches stochastically between assembly
and disassembly states making MTs highly dynamic structures
\cite{MK84a}. This property of MTs is referred to as dynamic
instability and it is a nonequilibrium
process.  It is generally believed that the instability starts from
the hydrolysis of guanosine triphosphate (GTP) bound to tubulin converting it
to guanosine diphosphate (GDP).   This reaction is
exothermic and releases $\sim 8 kT$ energy per reaction
\cite{WIS89}, i.e. approximately $0.22$ eV per molecule
\cite{EAHH89} at room temperature.  Here $k$ is the Boltzmann constant and $T$ is the
absolute temperature.   Since GDP-bound tubulin favors dissociation, a MT
enters the depolymerization phase as the advancing hydrolysis
reaches the growing end of a MT.   This dynamic phase transition is called a
catastrophe.  As a result, MTs rapidly disassemble releasing GDP-tubulin
in the solution where reverse hydrolysis takes place followed by a re-polymerization phase
of MTs called a rescue.   Therefore, MTs constantly
fluctuate between growth and shrinkage phases.

Several theoretical models have
been proposed for a macroscopic description of these processes using nonlinear
classical equations
\cite{HK82,BSM90,VOC02,DL93,Fey94,Dog95,HV96,Jobs97,DY98,BR99,HZ02,VCO05}. 
These phenomenological models provide good agreement with experiment but
generally consider a MT
as a one-dimensional mathematical object that switches stochastically
between growth and shrinkage states. Even two-dimensional \cite{HK82,BSM90,VOC02}
and three-dimensional \cite{VCO05} considerations for MT polymerization have started
from a given 2D (planar) or 3D (cylindrical) mathematical configuration that can grow or
shrink randomly.

In this paper, however, we based our model on fundamental
biochemical reactions that are occurring during microtubule's assembly
and disassembly processes.  This enables us to derive a nonlinear system of equations
that determines the structure, dynamics and the motion of MTs in three dimensions.
In particular, our model explains the continuum symmetry breaking of an isotropic pool of
tubulin dimers that leads to formation of experimentally observed
3D structures such as ring-shaped or filaments \cite{Ung90,Beh06,Hab06}.
We believe that this treatment is necessary to address the fundamental issues
about the observed dynamical behavior of MTs.  As stated by Del Giudice et al. \cite{Del85} :
``Systems with collective modes are naturally described by field theories. Furthermore,
quantum theory has proven to be the only successful tool for describing
atoms, molecules and their interactions."

\section{The Model}

Consider an individual MT in a free tubulin solution containing a large number
of GTP-tubulin, GDP-tubulin and a pool of free GTP molecules.  In
this solution several processes take place (as well as their reverse
reactions): \\
(i) GTP hydrolysis:
$$ \GTP \longrightarrow \GDP + \Delta_1,$$
(ii) conversion of tubulin GDP from tubulin GTP:
$$  \Tu_\GTP ~\longrightarrow~ \Tu_\GDP + \Delta_2,$$
(iii) growth of a MT:
$$\Delta_3 + \MT_{N-1} + \Tu_\GTP ~\longrightarrow~ \MT_N,$$
(iv) shrinkage of a MT:
$$ \MT_N  ~\longrightarrow~  \MT_{N-1} + \Tu_\GDP
+ \Delta_4.$$
Experimental studies determined the free energy values
for these reactions as: $\Delta_1 \simeq 220$ meV, $\Delta_2 \simeq
160$ meV and $\Delta_3 \simeq \Delta_4 \simeq 13\times 40 = 420$ meV, respectively
\cite{Cap94}.  
They are clearly above the thermal energy at room
temperature ($kT\simeq 26$ meV) and within a quantum mechanical
energy range that corresponds to the creation of one or a few
chemical bonds. Hence we may consider each chemical reaction as a
quantum mechanical process \cite{GTP}. As a result, an individual
microtubule with length $L$ can be viewed as consisting of $N$
tubulin layers defining its quantum state $|N\rangle$.
A tubulin layer consists of at least one tubulin dimer and at
most 13 tubulin dimers as observed in the MT's structure.
The state can
be raised/lowered by a creation/annihilation operator (i.e.
polymerization/depolymerization process) to the
$|N+1\rangle/|N-1\rangle$ state.  The corresponding MT is then
longer/shorter by one tubulin layer compared to the original one.

Further simplification will be achieved by combining the above processes into two
fundamental reactions:\\
(I) growth of a MT by one dimer length by adding of one tubulin layer in
an endothermic process:
\st
\be\label{I}
\Delta + \MT_{N-1} + \Tu_\GTP ~\longrightarrow~ \MT_N,
\ee
(II) shrinkage of a MT by one dimer length due to the
removal of one layer of $\Tu_\GDP$ dimers in an exothermic process:
\st
\be\label{II}
\MT_N  ~\longrightarrow~  \MT_{N-1} + \Tu_\GDP + \Delta,
\ee
where $\Delta$ is the energy of the reaction.
In order to derive a quantum mechanical description of mechanisms (\ref{I}) and (\ref{II}),
we introduce quantum states of a MT, tubulin and the heat bath, respectively:
\begin{itemize}
\item $|N\rangle$ is the state of a microtubule with $N$ dimers (containing
both GTP and GDP tubulins).
\item $|N_T\rangle$ is the state of a tubulin dimer, $\Tu_\GTP$ or $\Tu_\GDP$.
\item $|\tilde{N}\rangle$ is the GTP hydrolysis energy state.
\end{itemize}
Then, the relevant second quantization operators would be:
\st
\bea
\label{a_dagger}
&&a^\dagger=|N+1\rangle\langle N|,~~~~~~
a=|N-1\rangle\langle N|,\no\\
&&b^\dagger=|N_T+1\rangle\langle N_T|,~~~
b=|N_T-1\rangle\langle N_T|,\\
&&d^\dagger=|\tilde{N}+1\rangle\langle \tilde{N}|,~~~~~~
d=|\tilde{N}-1\rangle\langle \tilde{N}|.\no
\eea
Here $b$/$b^\dagger$ and $d$/$d^\dagger$ are annihilation/creation operators
of tubulin and energy quanta, respectively.  The operators $a$/$a^\dagger$ are
lowering/raising the number of tubulin layers in a MT.
The creation and annihilation operators obey the Bose-Einstein commutation relations
\st
\be
[ q_\k, q_\m^\dagger]=\delta_{\k \m},~~{\rm and}~~
[q_\k^\dagger,q_\m^\dagger]=0=[q_\k,q_\m],
\ee
where $[A,B]=AB-BA$ is the Dirac commutator or $q=a,b,~{\rm and}~ d$ .
Following \cite{TD01}, one can express the above processes using creation and
annihilation operators (\ref{a_dagger}):
\bea
\st\label{met1}
a^\dagger b ~d  &:& \Delta + \MT_{N-1} + \Tu_\GTP \longrightarrow  \MT_N \\
\st\label{met2}
d^\dagger ~ b^\dagger ~ a &:& \MT_N  \longrightarrow
\MT_{N-1} +  \Tu_\GDP + \Delta
\eea
Operators (\ref{met1}) and (\ref{met2}) describe a MT's growth and shrinkage
by one layer, respectively.  The polymerization or depolymerization
process may happen repeatedly before reversing the process which can be captured
by constructing product operators, i.e. $(a^\dagger b ~d)^m$ and
$(d^\dagger ~ b^\dagger ~ a)^n$, where $m$ and $n$ are the number of growing or shrinking events in
a sequence, respectively.
Based on the mechanisms in (\ref{met1}) and (\ref{met2}), the Hamiltonian
for interacting microtubules with $\Tu_\GTP/\Tu_\GDP$ tubulins can be written as
\st
\bea\label{Ham1}
H&=& \sum_\k \hbar\omega_\k a_\k^\dagger a_\k
+  \sum_\m \hbar\varpi_\m b_\m^\dagger b_\m
+  \sum_\l \hbar\sigma_\l~ d_\l^\dagger d_\l \no\\
&&\hspace{-1cm}+ \sum_{n=1}^\infty \sum_{\tilde{\k}_n, \tilde{\m}_n, \tilde{\l}_{n-1}}
  \hbar[ \Gamma_{\tilde{\k}_n \tilde{\m}_n \tilde{\l}_n} ~ c_{\tilde{\k}_n
  \tilde{\m}_n \tilde{\l}_n}
+   \Gamma^*_{\tilde{\k}_n \tilde{\m}_n \tilde{\l}_n } ~
c^\dagger_{\tilde{\k}_n \tilde{\m}_n \tilde{\l}_n} ],~~~~
\eea
where $\omega$, $\varpi$, $\sigma$ and $\Gamma$ are constants that can be related
to the energy of the fundamental processes \cite{DTC}.
A growing/shrinking MT may change its state
quickly or after several steps to a depolymerizing/polymerizing state and then
may change back to a polymerizing/depolymerizing state.
Experimentally, at a mesoscopic level the transition from the growing to the shrinking phase
is quantified by the catastrophe rate $f_{\rm cat}$ and the transition from the shrinking to
the growing phase is expressed by
the rescue rate $f_{\rm res}$ in which $f_{\rm res} < f_{\rm cat}$.
In the Hamiltonian (\ref{Ham1}), these transitions correspond to
a combination of creation and annihilation operators as the $n^{\rm th}$ power of
the reaction in
(\ref{met1}) and (\ref{met2}):
\st
\be\label{c-op}
c_{\tilde{\k}_n \tilde{\m}_n } =
(a^\dagger_{\k_1} b_{\m_1} ~d_{\l_1}) (a^\dagger_{\k_2} b_{\m_2} ~d_{\l_2})
\ldots
(a^\dagger_{\k_n} b_{\m_n} ~d_{\l_n }).
\ee
Here $\tilde{\k}_n=\{\k_1, \k_2, \ldots, \k_n\}$ is a collection of indices and
$\sum_{\tilde{\k}_n}=\sum_{\k_1} \sum_{\k_2} \ldots \sum_{\k_n}$.
We note that the momentum conservation for the last two terms in the
Hamiltonian (\ref{Ham1}) requires that
$\l_n = \sum_{i=1}^n \k_i - \sum_{i=1}^n \m_i - \sum_{i=1}^{n-1} \l_i$.
Therefore, the first $n-1$ of $\l$ will be free and summed in the Hamiltonian (\ref{Ham1}).

\section{The dynamical equations}
The Heisenberg equation of motion for a space- and time-dependent operator $q(\r,t)$
reads:
$i\hbar \partial_t{q}(\r,t) = - [H,q(\r,t)],$
where $H$ is the Hamiltonian.
A system of coupled equations that describes the quantum dynamics of a MT
can be derived from the Heisenberg equation. 
However, since MTs are overall classical objects
(although some of their degrees of freedom may behave as quantum
observables), we need to ensemble average over all possible states to
obtain effective dynamical equations.
Fourier transforming $a_\etas$, $b_\etas$ and $d_\etas$ operators
over all states, we find
\bea
\st
&&
\psi(\r,t) = \Omega^{-1/2} \sum_\etas \exp(-i \eeta\cdot\r) a_\etas (t),\\
\st
&&
\chi(\r,t) = \Omega^{-1/2} \sum_\etas \exp(-i \eeta\cdot\r) b_\etas (t),\\
\st
&&
\phi(\r,t) = \Omega^{-1/2} \sum_\etas \exp(-i \eeta\cdot\r) d_\etas (t),
\eea
where $\Omega$ is the volume over which the members of the plane wave basis are
normalized \cite{TD89a,DT95}.   Here $\psi(\r,t)$, $\chi(\r,t)$, and $\phi(\r,t)$
are the corresponding field operators for the quantum operators $a_\etas$, $b_\etas$ and
$d_\etas$, respectively.   The derivation of the equation of motion for the field
operators is lengthy but straightforward and given in detail in \cite{RT07}.
The final form of the equations of motion is found to be
\st
\bea\label{eqMT_s1}
&& i\partial_t{\psi} +  i\v\cdot \nab\psi =
-b \nabla^2 \psi +  V \psi, \\
\st\label{eq_chi1}
&&i\partial_t{\chi} =
-e \nabla^2 \chi + U \chi,\\
&& V(|\psi|,|\chi|) = a + c |\chi|^4 |\psi|^2 -d |\chi|^6 |\psi|^4,\no\\
&& U(|\psi|) = f - h |\psi|^2,\no
\eea
where $|\psi|^2=\psi\psi^*$.
\begin{center}
\begin{table}[th]
\caption{The parameters available in the literature. }\label{Tab1}
\begin{tabular}{lcll}
\hline
Parameter & Simulation Coeff.~~~ & ~~~~Exp. Value~~ & Reference\\\hline
MT growth rate  & Real($c$) & $0.50 - 19.7$ ($\mu$m/min)& \cite{PW02}\\
MT shortening rate   & Real($d$) &   $4.1 - 34.9  $ ($\mu$m/min)& \cite{PW02}\\
MT catastrophe frequency  & & $0.12 - 3.636$ (/min)& \cite{PW02}\\
MT diffusion constant   &  $b$ & $2.6-30.3$ ($\mu$m$^2$/min)& \cite{Maly02}\\
Tubulin diffusion constant ~~ & $e$ & $300-480$ ($\mu$m$^2$/min)& \cite{Odde97} \\
\end{tabular}
\end{table}
\end{center}
For simplicity we assumed that the energy in the system is distributed uniformly
during the course of experiment.  As a result, $\dot{\phi}=0$ and $\nab\phi={\bf 0}$.
Here parameters $a, b, e$ and $f$ are real but $c, d$ and $h$ are complex.
Table 1 lists experimental values for some of these parameters.
Eq. (\ref{eqMT_s1}) represents the
nonlinear cubic-quintic Schr\"{o}dinger (NLS) equation with a complex potential
that has been extensively studied in connection with topics such as
pattern formation, nonlinear optics, Bose-Einstein condensation, superfluidity
and superconductivity, etc.  \cite{AK02}.
A general solution of the NLS equation can be cast in the form of
\st
\be
\psi(\r,t)=R(\r,t)\exp[iS(\r,t)],
\ee
which involves topological defects (point in 2D and line in 3D).
In 3D space these defects represent 1D strings or vortex filaments \cite{AK02}.
Furthermore, it is shown that the symmetry group of the cubic-quintic NLS
equation is an extended Galilei group that includes translational and rotational symmetries
as well as proper Galilei boosts and total mass conservation \cite{GW89c}.
Adding any constrains such as boundary conditions to the equation, however,
will cause a symmetry breaking in the system.
As an example, in a cylindrical coordinate system, due to the rotational symmetry breaking,
the general solution reduces to a stationary solution
that represents a straight vortex filament with a twist:
\st
\be\label{sol}
\psi(r,\theta,z,t) = R(r) \exp[i(\omega t + n \theta + w(r) + k_z z)],
\ee
where $\omega$ is the spiral frequency,  $R(r)$ is the amplitude,
$w(r)$ is the spiral phase function and integer $n$ is the winding number of the vortex
\cite{GW89c,AB97}.
The axial wave number $k_z$ characterizes the vortex's twist.
In the case of the NLS equation, a family of vortices that move with a
constant velocity is also a solution \cite{AB97}.

Further symmetry breaking would lead to different 3D structures such as
double-wall , ring-shaped, sheet-like, C-shaped and S-shaped ribbons, and hoop structures
as seen during tubulin polymerization experiments \cite{Ung90,Beh06,Hab06}.

\section{Numerical results}
Equations (\ref{eqMT_s1}) and (\ref{eq_chi1}) are solved numerically with a no-flux boundary condition.
As an initial condition we chose a straight vortex filament perturbed by small noise (eg.
thermal or environmental noise).
In Fig. {\ref{fig:MT}
we compare the observed data on the MT length as a function of time with our simulation results.
The length of a vortex is defined as
\cite{AB97}:
\st
\be\label{L0}
L(t) = \int \Theta(\psi_0 - |\psi(\r, t)|) d^3r,
\ee
where $\Theta(x)$ is the step function and $\psi_0$ is a constant.
In Figs. {\ref{fig:MT}} and
\ref{fig:MT1} we compare the observed data on the MT length as a function of time with our simulation results.
Experimental panels in Figs. \ref{fig:MT} and \ref{fig:MT1}
represent the experimental data published by Rezania \emph{et al.} \cite{Rez08}.
Simulation panels show the numerical results of the normalized vortex length as a
function of time for the given set of
parameters. 

\begin{center}
\begin{figure}
\vspace{-.5cm}
\includegraphics[width=1\hsize]{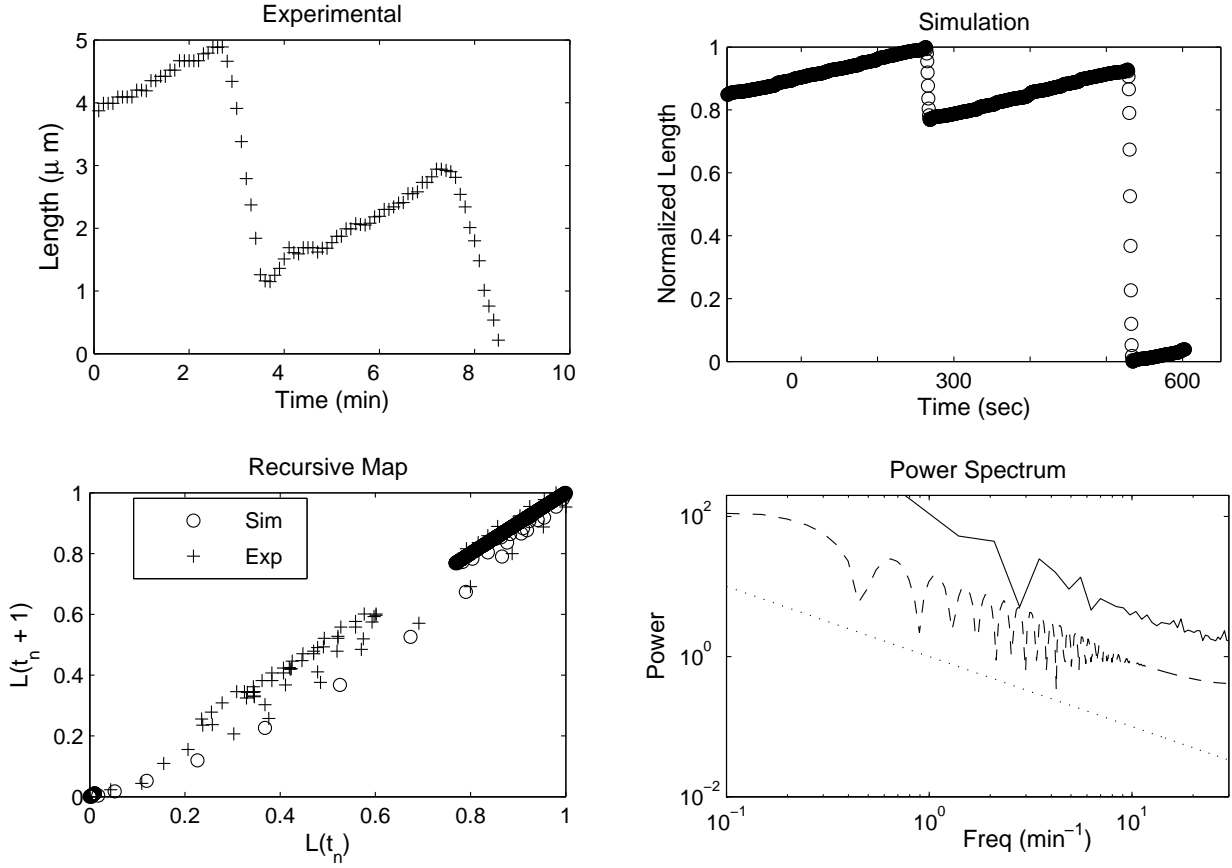}
\vspace{-.5cm} \caption{ Length of a distinct microtubule as a
function of time. The top-left panel represent experimental data
published in \cite{Rez08}.
provided by O.
Azarenko and M.A Jordan from the University of California, Santa
Barbara.
The top-right panel is the simulation result with the set of
parameters ($a=1, b = 10, c = 10 + i, d=20 + i, e =300, f =1$ and
$h= -.1+i$).  The bottom-left panel shows recursive maps for both
experimental and simulation results. The bottom-right panel
represents the power spectrum of the experimental (solid curve) and
simulated (dashed curve) data, respectively. The curves are plotted
at different offsets for clarity. No particular frequency of
oscillation can be seen from the power spectrums.  However, both
power spectrums show a very similar broad distribution that more or
less decays with frequency as an inverse power-law with slope $\sim
1.0$. The best fit inverse power-law is shown by a dotted line.
}\label{fig:MT}
\end{figure}
\end{center}
\begin{center}
\begin{figure}
\vspace{.5cm}
\includegraphics[width=1\hsize]{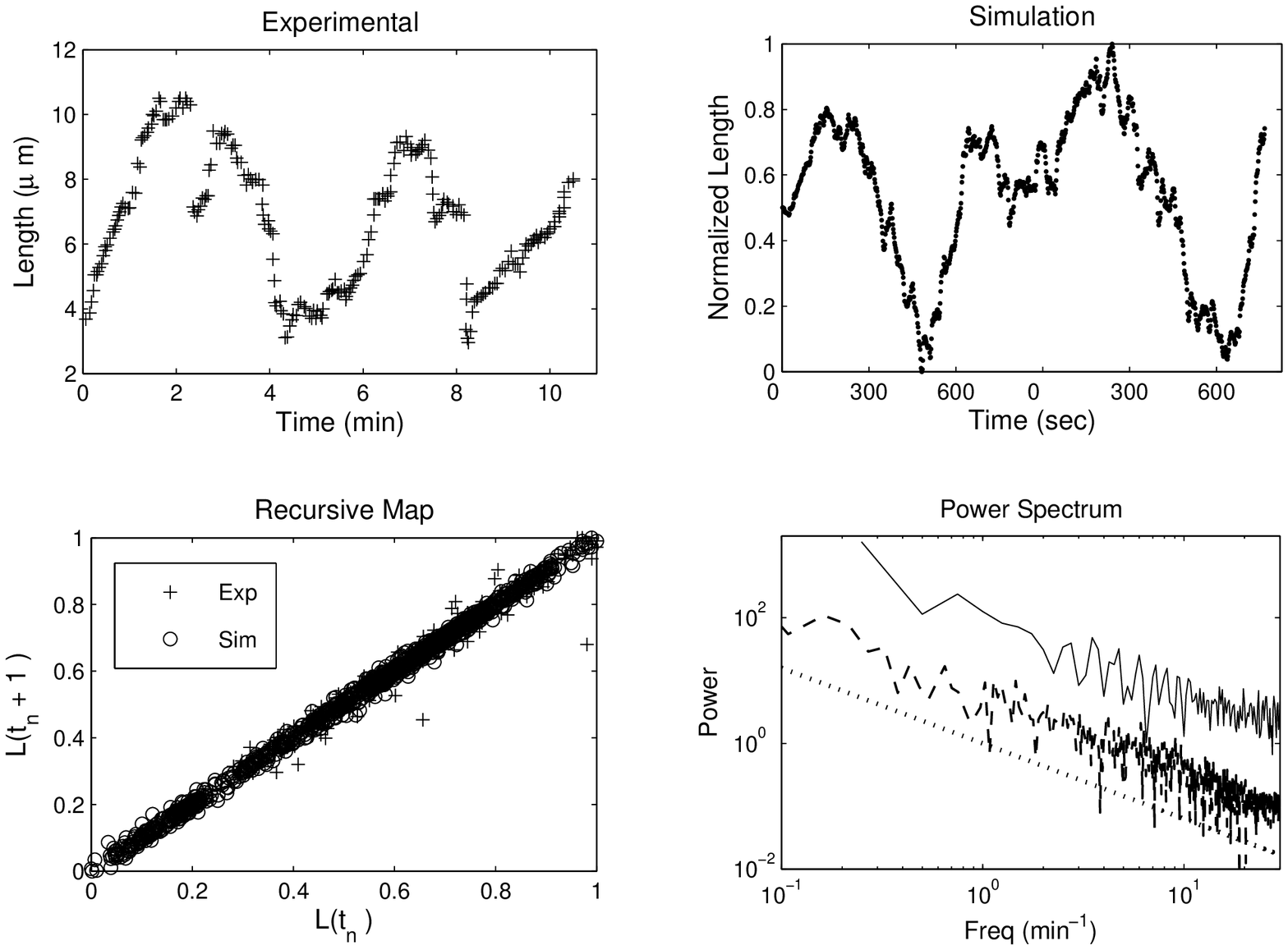}
\vspace{-.5cm} \caption{ Same as Fig 1. but with the set of
parameters ($a=1, b = 30, c = 10 + 10i, d=20 +10 i, e =300, f =1$
and $h= -.1+i$).  In the power spectrum panel, the inverse power-law
has a slope of $\sim 1.2$. }\label{fig:MT1}
\end{figure}
\end{center}
\vspace{-2.4cm}

To provide a simple yet accurate and powerful comparison between experimental and simulation results,
we graph recursive maps for the data points. The advantage of the recursive maps is the introduction of regularity into the data sets that allows for a better choice of adjustable parameters due to noise reduction inherent in the separation of data into subsets corresponding to independent processes.  In spite of being very simple, recursive maps of assembly and disassembly processes of individual MTs can successfully reproduce many of the key characteristic features. Consider first the following stochastic map as the simplest case that illustrates the approach taken:
\st
\be
\ell (t_n + 1) =  r[ \ell(t_n) + \delta],
\ee
where  $\ell(t_n)$ is the length of a microtubule after $n$  time steps, $t_n$.
The parameter $r$  is chosen to be a random number with the following two possibilities:
$$
r =  \left\{ \begin{array}{l}
      1 \;\;\; {\rm with \;probability}\;\; p\\
      0 \;\;\; {\rm with\; probability} \;\;1-p
      \end{array}
      \right.
      $$
In terms of the MT polymerization process, $p$ is the probability
that a given event will result in assembly while $1-p$ is the
probability of a complete catastrophe of the MT structure. The above
simplified model, therefore, is governed by only two adjustable
parameters: (a) the probability of complete catastrophe $1-p$ which
is constant and independent of the length or time elapsed and (b)
the rate of polymerization which is proportional to the length
increment $\delta$ over the unit of time chosen in the simulation.
Thus, the coefficient $\delta$ divided by the time step $\Delta t$
($ = t_{n+1} - t_n$) gives the average  growth velocity of an
individual MT. Such information can be used to fine tune the
simulation parameters. We note that the slope of the line in the
simulation panels in Fig. \ref{fig:MT} can be adjusted by varying
the real parts of parameters $c$ and $d$.  The frequency of
catastrophe events can also be changed by adjusting the parameter
$b$. In the recursive map panels in Figs. \ref{fig:MT} and
\ref{fig:MT1} we compare the recursive map for both the experimental
data and the simulation results. Based on the recursive maps, the
key characteristics of the experimental and simulated results that
were obtained independently are quite similar.   This represents
Eqs. (\ref{eqMT_s1}) and truly describes the dynamics of MTs'
polymerization.

To provide a more solid comparison, a spectral analysis is also
carried on both experimental and simulated data. As discussed by
Odde \emph{et al.} \cite{OBC96}, the power spectrum analysis is a
more general way to characterize the microtubule
assembly/disassembly dynamics without assuming any model \emph{a
priori}.  The power spectrum panels in Figs. \ref{fig:MT} and
\ref{fig:MT1} represent the spectral power of the experimental and
simulated data, respectively. As shown, there is a great agreement
between the experimental (solid curve) and simulated (dashed curve)
spectrums. We note that the curves are plotted at different offsets
for visual clarity only. As expected, no particular frequency of
oscillation can be found from the power spectrums.  However, both
power spectrums demonstrate a very similar broad distribution that
more or less decays with frequency as an inverse power-law with
slope $\sim 1.0$ in Fig. \ref{fig:MT} and $\sim 1.2$ in Fig.
\ref{fig:MT1}, respectively. The best fit inverse power-law is shown
by a dotted line in both panels.

Although this is not unique to the model presented here,
our results show no attenuation states during MT polymerization
(Figs. \ref{fig:MT} and \ref{fig:MT1}).  The MT length undergoes small fluctuations
all the time. This
can be understood by noting that our model is based on the cyclic polymerization and
depolymerization of tubulin dimers.  Behavior consistent with this result has recently
been observed
by Schek \emph{et al.}\cite{Sch07} who studied the microtubule assembly dynamics
at higher spatial ($\sim$ 1-5 nm) and temporal ($\sim$ 5 kHz) resolutions.
They found that even in the growth phase, a MT undergoes shortening excursions at the
nanometer scale.

\section{Discussion}
The basic structural unit of a MT is the tubulin dimer. Each dimer exists in a quantum
mechanical state characterized by several variables, especially GTP/GDP.
Each microstate of a tubulin dimer is sensitive to the states of its neighbors.
Tubulin dimers have both discrete degrees of freedom (distribution of charge) and
continuous degrees of freedom (orientation). A model that focuses on the discrete
variables will be an array of coupled binary switches \cite{Ras90}, 
while a model that focuses on the continuous ones could be an array of coupled
oscillators \cite{BT97,Sam92}.  Here we have focused on tubulin binding and
GTP hydrolysis as the key biochemical processes determining
the states of MTs which are most easily
accessible to experimental determination.  We have shown how a quantum
mechanical description of the energy binding reactions taking place during MT polymerization
can lead to nonlinear field dynamics with very rich behavior that includes both localized
energy transfer and oscillatory solutions.
%

Based on the chemical binding reactions occurring during MT
polymerization, a quantum mechanical Hamiltonian for the system has been proposed
in this work.
Equations of motion have then been derived and transformed from the purely quantum mechanical
description to a semi-classical picture using the method of coherent
structures. We found that the dynamics of a MT can be explained by the
cubic-quintic nonlinear
Schr\"{o}dinger equation (NLS) with a complex potential.  A generic solution
of the NLS equation in cylindrical
geometry is a vortex filament \cite{AK02,GW89c,AB97} which
can grow or shrink linearly in time as well as fluctuate temporally with some frequency.
This behavior exhibits two distinct dynamical phases: (a) linear
growth/shrinkage and (b) oscillation about a mean value, which are indeed main experimentally observed
characteristics of the MT dynamics (Fig. 1 and 2).\\

We have demonstrated here that the
assembly process can be described starting from quantum mechanical
first principles applied to biochemical
reactions.   This can be subsequently transformed into a highly nonlinear
semi-classical dynamics problem.  The gross features of MT dynamics satisfy
classical field equations in a coarse-grained picture.  Individual chemical
reactions involving the constituent molecules still retain their quantum
character.  The method of coherent structures allows for a simultaneous
classical representation of the field variables and a quantum approach to their
fluctuations.   Here, the overall MT structure (and their ensembles) can be
viewed as a virtual classical object in (3+1)-D space-time.   However,
at the fundamental level of its constituent biomolecules, it is quantized as
are chemical reactions involving its assembly or disassembly.
In our approach the route taken is opposite since we
started with individual tubulin quantum microstates to arrive at classical,
nonlinear but classically coherent (and stable) macro-states of a microtubule.

Finally, we note that dynamics of pattern formation can be also described by NLS equation in which
$\psi(\r,t)$ represents the order parameter.  A number of intriguing experiments
performed by Tabony \emph{et al.} \cite{Tab}
demonstrated that gravity can influence tubulin assembly reactions with MTs forming
distinct bands at right angles to the orientation of the gravity field or, if spun, to the centrifugal force.
Despite several studies \cite{Por03,Por05,AT05,AT06}, the above experiments are yet to be fully explained theoretically.
Our goal in future studies is to
focus on the dynamics of pattern formation by MTs using the results presented in this paper.

\emph{Acknowledgment} This research was supported in part by the Natural Sciences and
Engineering Research Council of Canada (NSERC) and the Canadian Space Agency (CSA).
Insightful discussions with S. R. Hameroff and  J. M. Dixon are gratefully
acknowledged.  The authors would also thank  M. A. Jordan for providing microtubule assembly data.
VR specially thanks I. Aranson for sharing his CGLE code and for fruitful discussions.

----------------------------------------------------------------------------------------


\begin{thebibliography}{59}
\bibitem{GTP}
We note that GTP hydrolysis is a subtle biochemical
process that carries a quantum of biological energy and thus allows us
to make a link between quantum mechanics and polymer dynamics.
\end{thebibliography}
\end{document}